\input harvmac
\input epsf

\def\frak#1#2{{\textstyle{{#1}\over{#2}}}}
\def\frakk#1#2{{{#1}\over{#2}}}

\def \in{\leftskip = 40 pt\rightskip = 40pt}
\def \out{\leftskip = 0 pt\rightskip = 0pt}

\def\ga{\gamma}

\def\sy{supersymmetry}
\def\sic{supersymmetric}

\def\cmp{Comm.\ Math.\ Phys.\ }

\def\npb{{Nucl.\ Phys.\ }{\bf B}}

\def\plb{{Phys.\ Lett.\ }{\bf B}}

\def\sjnp{Sov.\ J.\ Nucl.\ Phys.\ }
\def\tmp{Theor.\ Math.\ Phys.\ }

{\nopagenumbers
\line{\hfil LTH 390}
\line{\hfil hep-ph/9702304}
\line{\hfil Revised Version}
\vskip .5in
\centerline{\titlefont Large-$N$ supersymmetric $\beta$-functions}
\vskip 1in
\centerline{\bf P.M.~Ferreira, I.~Jack and D.R.T.~Jones}
\bigskip
\centerline{\it Dept of Mathematical Sciences, 
University of Liverpool, Liverpool L69 3BX, U.K.}
\vskip .3in

We present calculations of the leading and $O(1/N)$ terms in  a large-$N$ 
expansion of the $\beta$-functions for various \sic\ theories:  a
Wess-Zumino model, \sic\ QED and a non-abelian \sic\ gauge theory. 
In all cases $N$ is the number of a class of the chiral superfields in 
the theory. 

\Date{Feb 1996}

Coupling constant perturbation theory has constituted  the main approach
to quantum field theories since their introduction, but it has its 
limitations. Therefore any approach which reaches beyond it is  worthy
of attention: one such is large-$N$ expansions, where $N$ denotes the 
number of fields (or some subset thereof). Simple theories such as 
$O(N)$-symmetric $\phi^4$ have been much studied
\ref\eya{M.~Campostrini and P.~Rossi, 
Int. J. Mod. Phys. {\bf A}7 (1992) 3265\semi 
G.~Eyal et al, \npb 470 (1996) 369\semi
D.J.~Broadhurst, J.A.~Gracey and D.~Kreimer, hep-th/9607174, Z. Phys. C in 
press} as has QCD at both large $N_c$\ref\witt{ E. Witten,  Ann. Phys. 128 (1980) 363} 
and $N_f$\ref\gracAeya{J.A.~Gracey, Int. J. Mod. Phys. {\bf A}8 (1993) 2465; 
\plb 373 (1996) 173}
\ref\graceyb{ J.A.~Gracey, \npb 414 (1994) 614}. In this paper we 
calculate the $\beta$-functions for a Wess-Zumino model, and also  for
various supersymmetric gauge theories--
\sic\ QED in the limit of large $N_f$, i.e. for a large number  of
chiral superfields, together with an abelian gauged Wess-Zumino model and 
a more general non-abelian 
theory. In all cases the leading contribution  is a simple
one-loop calculation, and things become interesting at $O(1/N)$, when
bubble sums are involved. It turns out that the  \sic\ D-algebra part of
the calculation is quite straightforward;  the resulting Feynman
integral calculation appears formidable,  but simplifies in miraculous
fashion, as observed for similar  (non-\sic) calculations in 
Ref.~\ref\pmpp{A. Palanques-Mestre  and P.~Pascual, \cmp 95 (1984) 277}.
\newsec{The Bubble Sums}

In this section we describe the Feynman integral calculations. We 
do all calculations with zero external momentum, using 
\sic\ dimensional regularisation (with $d = 4-2\epsilon$)  and minimal
subtraction (DRED). By performing subtractions at the  level of the Feynman
integrals we completely separate  the calculation of the (subtracted)
Feynman integrals from  the details of the theory under consideration. 
It is convenient to redefine the  $d$-dimensional integration
measure so that 

\eqn\bubba{
\int \frakk{d^d k}{k^2 (k-p)^2} = \pi^2\frakk{1}{\epsilon}(p^2)^{-\epsilon}.
}
Three diagrams of the kind we will require are shown in Figure~1; these 
will in fact suffice to derive all the results we present, except for those 
in section 5.  
Let us consider Fig.~1(B).  After subtracting all sub-divergences, 
the $n$-bubble (i.e. $(n+1)$-loop) 
contribution to this diagram is given by the expression: 
\eqn\bubbb{
B_n = \frakk{\kappa^n}{\epsilon^{n+1}} G(\epsilon)
\sum_{r=1}^{n+1} r^{-1}(1-r\epsilon)
\Gamma (1 + r\epsilon)\Gamma(1-r\epsilon)\pmatrix{n\cr r-1\cr}(-1)^{r+1} 
x^{r\epsilon}}
where
\eqn\gfunc{
G(\epsilon) = \frakk{\Gamma (2 -2\epsilon)}{\Gamma(2-\epsilon)
\Gamma(1-\epsilon)^2\Gamma(1+\epsilon)}} 
and $x = \mu^2$, $\mu$ being the regulator mass. 
The parameter $\kappa$ subsumes any constant factors 
which will recur on a bubble-by-bubble basis.  

\bigskip
\epsfysize= 1.0in
\centerline{\epsfbox{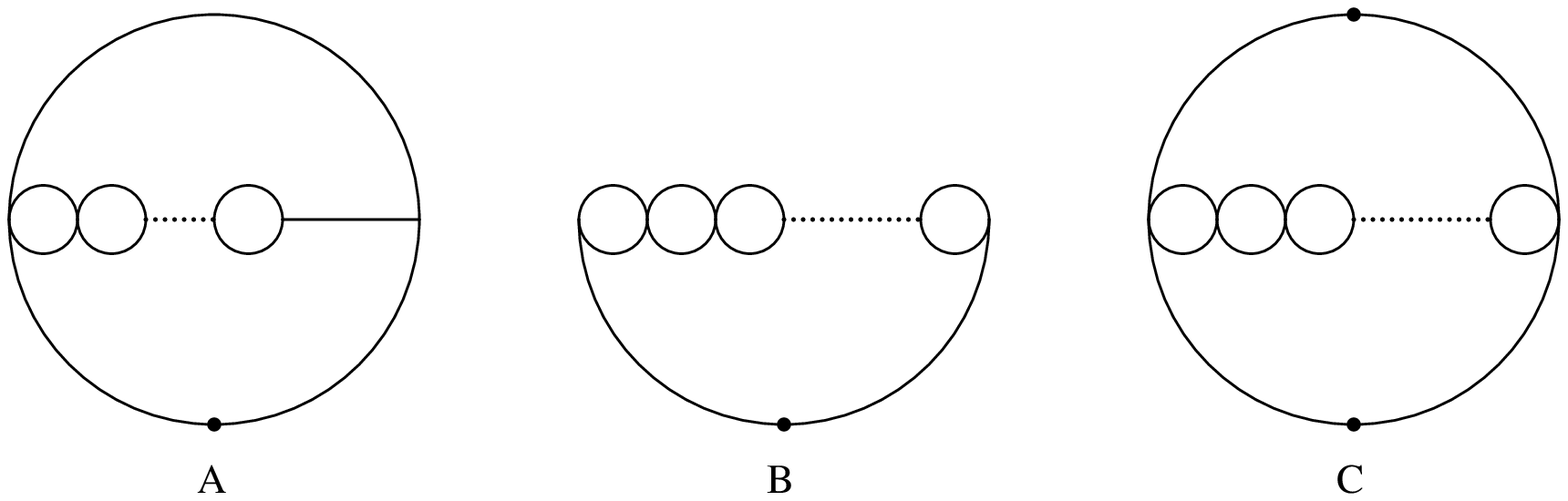}}
\in
{\it \noindent Fig.1:
Feynman diagrams representing the bubble sums $A, B$ and $C$. The black 
dots denote squared propagators}
\medskip
\out

We now write\pmpp
\eqn\bubbc{
(1-r\epsilon)
\Gamma (1 + r\epsilon)\Gamma(1-r\epsilon)
x^{r\epsilon} = \sum_{j=0}^{\infty}L_j (r\epsilon)^j.}
Substituting in Eq.~\bubbb, and using the identity
\eqn\bubbd{\eqalign{\Delta_j = 
\sum_{r=1}^{n+1}r^{j-1}\pmatrix{n\cr r-1\cr}(-1)^r
&= 0 \quad\hbox{when}\quad j = 1,2,\cdots n\cr
&= -(n+1)^{-1} \quad\hbox{when}\quad j = 0\cr}} 
we find that the pole terms in $B_n$ are given by the expression
\eqn\bubbe{
B_n^{\rm{pole}} = \frakk{\kappa^n}{(n+1)\epsilon^{n+1}}
\sum_{i=0}^n G_i \epsilon^i
}
where we have written $G (\epsilon) = \sum G_n \epsilon^n$.    
The identity Eq.~\bubbd\ removes all the non-local (i.e. $\ln x$-dependent) 
counter-terms.  Now we want to sum over $n$. 
In a $\beta$-function or anomalous dimension calculation, 
the result will be given by the coefficient of the simple pole in $\epsilon$ 
in the quantity $\sum (n+1)B_n^{\rm{pole}}$, which is easily seen to give
\eqn\bubbf{
B =\sum_{n=0}^{\infty}G_n\kappa^n = G(\kappa).}
Similar calculations give:
\eqn\bubbg{
A = - \kappa^{-1}\left[G(\kappa) -1 + 2\int_0^\kappa G(x)\,dx\right]}
and 
\eqn\bubbh{
C = - 2\kappa^{-1}\left[G(\kappa) -1 + \int_0^\kappa (1+2x)G(x)\,dx\right].}

Thus all the bubble sums relevant to our calculations depend on the function 
$G(x)$, which has a zero at $x=1$ and a simple pole at $x=\frak{3}{2}$.   
We may therefore anticipate that our results in subsequent sections 
will have a finite radius of convergence in the appropriate coupling 
constant, because of this pole. We turn now to explicit models.   

\newsec{The large-N Wess-Zumino model}

The superpotential of the model is 
\eqn\none{
W = \frakk{\lambda}{\sqrt{N}}\sum_{i=1}^{N}\phi\xi_i\chi_i.}
At leading order, it is trivial to see that $\beta_{\lambda}$ is 
determined by the one loop contribution to $\gamma_{\phi}$. 
The Feynman diagrams contributing  at $O(1/N)$ are shown in Fig.~2. 
\bigskip
\epsfysize= 1.0in
\centerline{\epsfbox{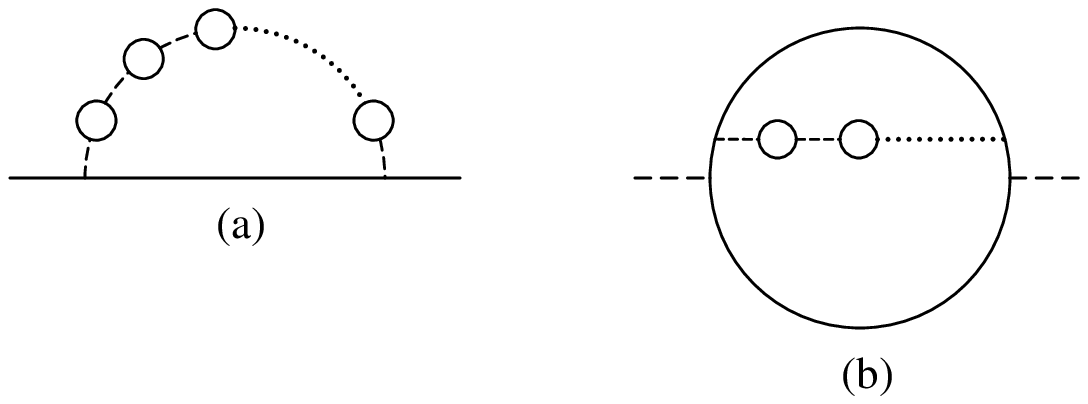}}
\in
{\it \noindent Fig.2:
The Feynman diagrams for section~2. Dashed lines are $\phi$-propagators, 
and solid lines are $\xi$ or $\chi$ propagators.}
\medskip
\out

We find 
\eqn\noneb{\eqalign{
\gamma_{\xi} &= \gamma_{\chi} = \frakk{y}{N}B(y) = \frakk{1}{N} y G(y),\cr
\gamma_{\phi} &= y\left[ 1 + 2\frakk{y}{N}A(y)\right] =
 y +\frakk{2y}{N}\left[1 - G(y) -
2\int_{0}^{y}G(x)\,dx\right].\cr}}
where
$y = \lambda^2/16\pi^2$. These results (and all our subsequent results for 
$\beta$-functions and $\gamma$-functions) are correct to $O(1/N)$.
Our result for $\beta_{\lambda} = 
\lambda\left[2\gamma_{\xi} + \gamma_{\phi}\right]$ is  thus 
\eqn\nbeta{
\beta_{\lambda} = \lambda y\left[1 + 
 \frakk{2}{N}H(y)\right]}
where
\eqn\nbetab{
H(y) = 1 - 2\int_0^{y}G(x)\, dx.}
It is quite straightforward to verify that Eq.~\nbeta\ reproduces the 
relevant terms in the existing four-loop calculation
\ref\fjj{P.M.~Ferreira, I.~Jack, and D.R.T.~Jones, \plb 392 (1997) 376}\ for a 
generalised Wess-Zumino model. 
\newsec{Supersymmetric QED}

In this section we consider \sic\ QED with $M=2N$  charged chiral
superfields $\xi, \chi$,  with pairs of charges  $\pm g/\sqrt{N}$,  for
large $N$. As for the WZ model,  the dominant contribution to $\beta_g$
is one-loop. The  graphs for the $O(1/N)$ calculation are shown in
Fig.~3. 

\bigskip
\epsfysize= 1.5in
\centerline{\epsfbox{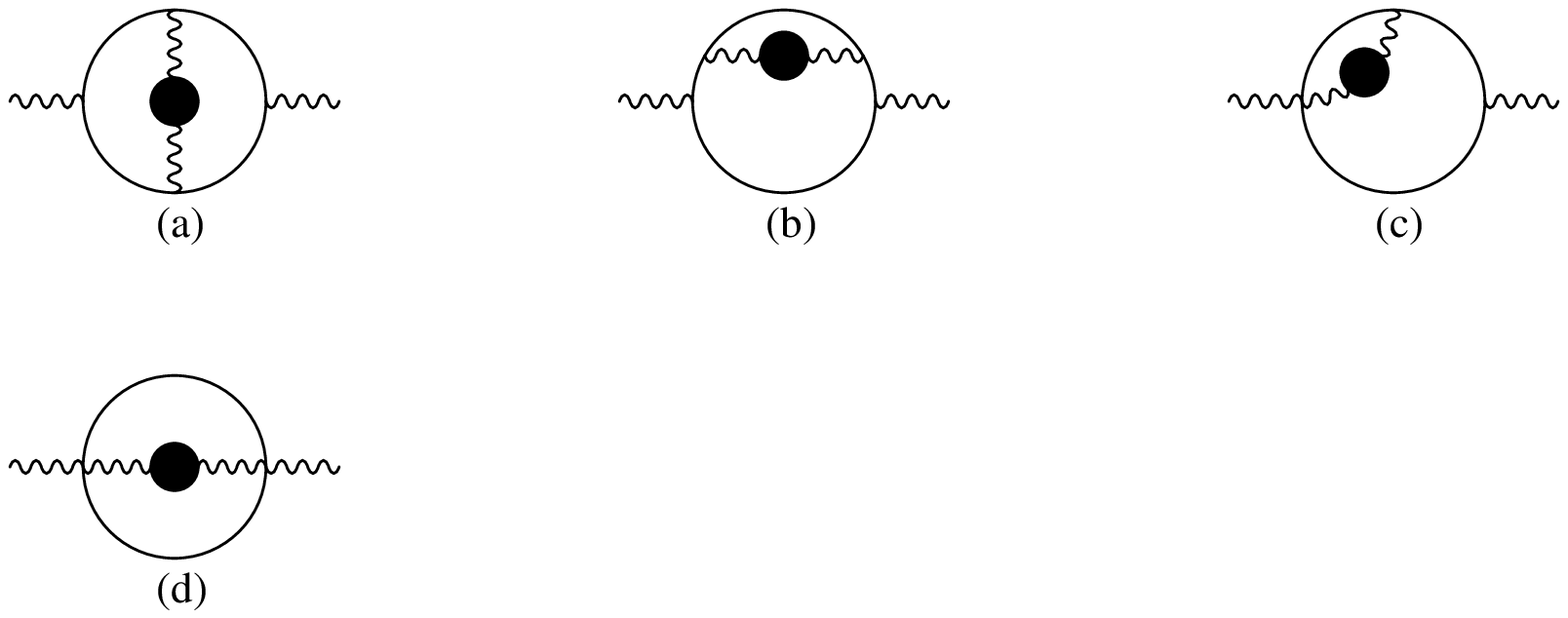}}
\in
{\it \noindent Fig.3:
Feynman diagrams for section~3. Wavy lines are vector propagators, 
and solid lines are $\xi$ or $\chi$ propagators. Blobs denote a chain 
of $\xi$ or $\chi$ bubbles.}

\medskip
\out

In Refs.~\ref\cjja
{I.~Jack, D.R.T.~Jones and C.G.~North, \plb 386 (1996) 138}, 
\ref\cjjb{I.~Jack, D.R.T.~Jones and C.G.~North, hep-ph/9609325, 
Nucl. Phys. B in press} we found $\beta_g$ for 
an abelian theory to four loops, by calculating the 
vector superfield self-energy in the Feynman gauge (note that in this gauge  
this suffices in the abelian case).
For details of our technique for dealing with the D-algebra part 
of the calculation we refer the reader to Ref.~\cjjb; the upshot is that 
we have simply to replace the Feynman integrals $A, B, C$ of that reference 
with the corresponding bubble summed quantities $A, B, C$ from this one.   
Our result for $\beta_g$ is 
\eqn\qeda{
\beta_g = gK\left[1 + \frakk{2}{N}\int_0^K (1-2x)G(x)\,dx\right]}
where $K = g^2/8\pi^2$, while for the anomalous dimension of each 
chiral superfield, $\gamma(g)$, we obtain
\eqn\qedb{
\gamma(g) = -\frakk{K}{N}G(K).}

It is interesting at this point to compare these results with 
the NSVZ all orders formula
\ref\nov{
V.~Novikov et al, \npb 229 (1983) 381\semi
V.~Novikov et al, \plb166 (1986) 329\semi 
M.~Shifman and A.~Vainstein, \npb 277 (1986) 456\semi
A.~Vainstein, V.~Zakharov and M.~Shifman,
\sjnp 43 (1986) 1028\semi
M.~Shifman, A.~Vainstein and V.~Zakharov \plb 166 (1986) 334}
for $\beta_g$, which for our theory 
reads:
\eqn\qedx{\beta_g^{NSVZ} =
gK\left[1- 2
\ga^{NSVZ}\right].}
We see that our results for $\beta_g$ and $\gamma(g)$ do not satisfy 
this relation. This is not surprising, because it was shown explicitly in 
Ref.~\cjja\ that the DRED and NSVZ $\beta$-functions part company at three 
loops. It is straightforward to construct order by order in $g$ the 
coupling constant redefinition that connects the two schemes. 
Because in this explicit example we have no non-trivial tensor structure, 
interesting constraints on the nature of the redefinition of the kind 
exploited in Refs.~\cjja, \cjjb\ do not occur.  
  
\newsec{General Abelian Theory}

Here we present results for a general theory produced by a $U_1$ gauging of 
the model defined by Eq.~\none. It is easy to see that for $N>1$ 
the constraints 
of gauge invariance of $W$ and anomaly cancellation mean that 
the most general 
gauging such that $q_{\chi_i}$ and $q_{\xi_i}$ are independent of $i$ is 
given by $q_{\phi} = 0$ and $q_{\xi} = - q_{\chi} = q$; we will set $q=1$.  
For the 
special case $\lambda = \sqrt{2}g$ we have 
${\cal N} =2$ \sy. In Fig.~4 we show the new Feynman diagrams we require for 
$\beta_g$, beyond those calculated in the previous section. 

\bigskip
\epsfysize= 1.0in
\centerline{\epsfbox{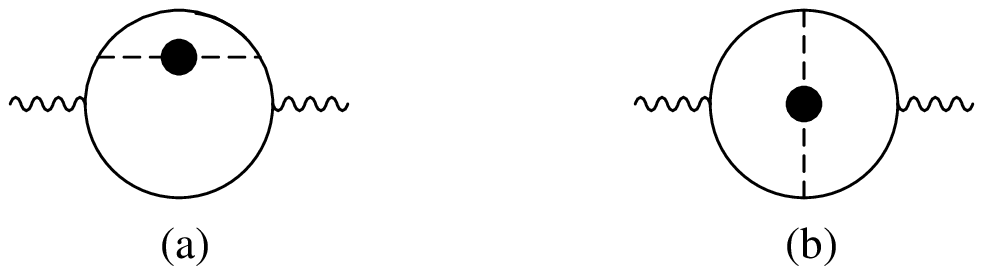}}
\in
{\it \noindent Fig.4:
Additional Feynman diagrams for section~4. Blobs denote a chain
of $\xi, \chi$ bubbles.}
\medskip
\out

The result is 
\eqn\gata{
\beta_g = gK\left[1 + \frakk{2}{N}
\int^{K}_{y} (1-2x)G(x)\,dx\right].}
We also find 
\eqn\gatd{\eqalign{
\gamma_{\xi}=\gamma_{\chi}&=\frakk{1}{N}\left[yG(y)-KG(K)\right],\cr
\gamma_{\phi}&= y+\frakk{2y}{N}\left[G(K)-G(y)+2\int_y^{K}G(x)\, dx
\right].\cr}}

It is again easy to verify that our result agrees with the three and four 
loop calculations presented in Ref.~\cjjb. Moreover, for ${\cal N}=2$ (which 
corresponds to $y = K$) we have $\beta_g = \gamma_{\phi}= 0$ beyond one loop, 
and $\gamma_{\xi} = \gamma_{\chi} = 0$ to all orders, in accordance with 
Ref.~\ref\hsw{
P.S.~Howe, K.S.~Stelle and P.~West, \plb 124 (1983) 55\semi
P.S.~Howe, K.S.~Stelle and P.K.~Townsend, \npb  236 (1984) 125}.

\newsec{General Non-Abelian Theory}
We now consider a non-abelian theory with gauge group ${\cal G}$ and 
superpotential
\eqn\nata{W=\frakk{\lambda}{\sqrt N}\phi^a\sum\xi_i^TS_a\chi_i,}
where $\xi_i, \chi_i, \phi$ are multiplets transforming under the 
$S, S^*$ and adjoint representations of ${\cal G}$ respectively. For 
notational simplicity we 
take the representation $S$ to be irreducible.  
In addition to diagrams
similar in form to those computed earlier in the abelian case, the two-point
function for the vector superfield 
includes the additional diagrams depicted in Fig.~5, because the 
$\phi$ field now has gauge interactions, as well as further 
diagrams involving the gauge coupling $g$ only. 
\bigskip
\epsfysize= 1.0in
\centerline{\epsfbox{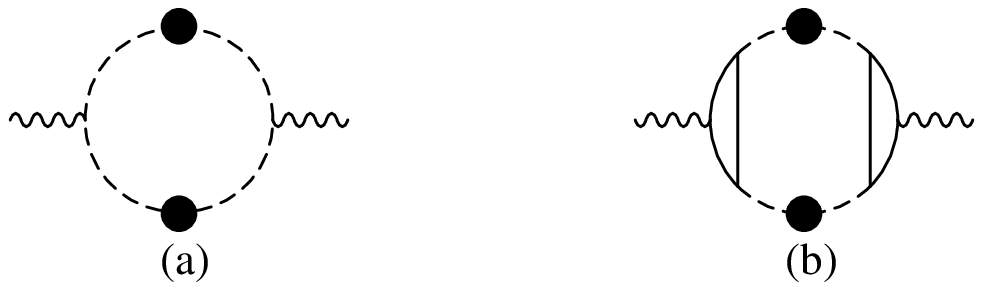}}
\in
{\it \noindent Fig.5:
Additional Feynman diagrams for section~5.}

\medskip
\out

The diagrams in Fig.~5 
give rise to 
bubble sums similar to $A,B$ and $C$ calculated earlier. 
The fact that these diagrams 
do not contain vector superfield propagators suggests that they 
correctly determine the corresponding contributions in the non-abelian case. 
(Note that the graph similar to Fig.~5(b) but with only one 
$\chi$ or $\xi$ loop gives no simple pole.) 
We can then infer the non-abelian result by using the afore-mentioned 
fact that there are no divergences 
beyond one loop  for ${\cal N}=2$. The result is

\eqn\natb{\eqalign{\beta_g=&gK\left[T(S)+\frakk{2\tr[C(S)^2]}{rNT(S)}
\int_{\hat y}^{\hat K}
(1-2x)G(x)\,dx\right]
\cr
\qquad&+\frakk{gK}{N}\left[\int_{\hat y}^{\hat K}G(x)\,dx -1
\right]C(G).\cr}}
For the chiral superfield anomalous dimensions we find: 
\eqn\natc{\eqalign{
\gamma_{\xi} &= \gamma_{\chi} 
= \frakk{1}{N}\left[yG(\hat y)-KG(\hat K)\right]C(S),\cr
\gamma_{\phi} &= \hat y+\frakk{2y\tr[C(S)^2]}{rNT(S)}\left[G(\hat K)-G(\hat y)+
2\int_{\hat y}^{\hat K}G(x)\,dx\right]\cr
&+\frakk{1}{N}(y - K) G(\hat K) C(G)-\frakk{y}{N}C(G)\cr}}
where $\hat y = yT(S)$ and $\hat K = KT(S)$. For definitions of the 
(fairly standard) group 
theory factors  $C(S)$, $T(S)$ and $C(G)$ see for instance Ref.~\cjja; $r$ is 
the number of generators of the group. 
The above  results contain as special cases all those presented in 
previous sections. Once again one can check compatibility with the three 
and four-loop calculations from Ref.~\cjjb\ and 
Ref.~\ref\cjjc{I.~Jack, D.R.T~Jones and C.G.~North, 
\npb 473 (1996) 308}.

\newsec{Discussion}

In a recent paper\fjj, we argued that the 
$\beta$-functions in simple  Wess-Zumino models (such as the
single field case) suggest that at $L$ loops one has 
$\beta^L \sim (-1)^{L+1}L!$ behaviour. Hence we suggested 
that they  are susceptible to Pad\'e-Borel
summation, and we argued that this was favourable  for the quasi-infra-red
fixed point scenario. The large $N$ regime dealt with here is clearly quite 
different in that we find that through $O(1/N)$ we have a finite radius 
of convergence in the coupling constant(s), caused 
by the pole at $x=\frak{3}{2}$ in $G(x)$.
 
It would clearly be interesting to see whether the 
finite radius of convergence mentioned above persists at  
higher orders in $1/N$. One might well expect, in fact, 
the $O(1/N^2)$ term to depend on $G^2$, or some convolution thereof. 
Perhaps the critical methods of
Ref.~\ref\honk{A.N~Vasil'ev, Yu.M. Pis'mak and J.R. Honkonen,  \tmp 46
(1981) 157; {\it ibid\/} 47 (1981) 291} could facilitate such calculations, 
as they have in the non-\sic\ case. 

Since the above calculations include contributions to all orders in 
perturbation theory, it behoves us, more than usual, to consider 
the issue of the potential ambiguities in DRED raised in 
Refs.\ref\siegelb,{W.~Siegel, \plb 94 (1980) 37},  
\ref\vlad{L.V.~Avdeev, G.A.~Chochia and A.A.~Vladimirov, \plb 105 (1981) 272}. 
The central tenets of these papers have not, to our knowledge, been 
challenged; and yet DRED has remained, by and large, the regularisation 
of choice for higher order \sic\ calculations.  

Although we do here include all orders of perturbation theory, 
the DRED ambiguities of \siegelb, \vlad\ do not, in fact, arise 
because of the ``bubble chain'' structure of the graphs. This would 
suggest that our calculation of the $O(1/N)$ contribution is well defined, but  
may not seem entirely satisfactory, since if there 
are ambiguous contributions  at any order of $1/N$ one may question the 
consistency of the regulator and the significance of the results. 
(This objection  could, of course, also be made 
to any of the many conventional perturbative DRED computations.)
We believe, however, that it should be possible to formulate 
the DRED ambiguities in a way which demonstrates them to be equivalent to 
scheme dependence ambiguities. In support of this conjecture, consider 
\ref\tim{D.R.T.~Jones, \plb 192 (1987) 391},
\ref\allen{R.W.~Allen and D.R.T.~Jones, \npb 303 (1988) 271}, which 
dealt with the metric and torsion $\beta$-functions for two-dimensional 
\sic\ $\sigma$-models.  
In \allen\ it was explicitly verified that 
if one requires the two-dimensional alternating tensor 
$\epsilon^{\mu\nu}$ to satisfy the equation:
\foot{In fact  relations of this type  
were first explored in the non-\sic\ context: 
see \ref\mtht{C.M.~Hull and P.K.~Townsend,  \plb 191 (1987) 115\semi
R.R.~Metsaev and A.A.~Tseytlin, \plb 191 (1987) 354\semi
I.~Jack and  D.R.T.~Jones, \plb 200 (1988) 453}.} 
\eqn\robina{
\epsilon^{\mu}{}_{\nu}\epsilon^{\nu\rho} = (1 + c\epsilon)g^{\mu\rho}}
(where here $\epsilon = 2-d$) then although the $\beta$-functions do depend 
on $c$ at two loops, this dependence can be removed by field redefinitions. 
This $c$-dependence is associated with a two-dimensional version 
of the ambiguity noted 
(in the four-dimensional case) by Siegel\siegelb;
we conjecture, therefore, 
that the four dimensional case may be dealt with in a similar way, 
with coupling constant redefinitions instead of  field redefinitions.

We hope to flesh out this idea, and also  consider applications 
of our results, in future publications.
  
\bigskip\centerline{{\bf Acknowledgements}}\nobreak

PF was supported by a  scholarship from JNICT. We thank John Gracey for
several useful  discussions. 

\listrefs 

\bye